\documentclass[cameraready]{Interspeech}
\usepackage{subfigure}

\title{A study on the impact of region specific data on the performance of Indic ASR}

\author[affiliation={1}]{Agneedh}{Basu}
\author[affiliation={1}]{Pavan K}{J}
\author[affiliation={1}]{Pranav}{Bhat}
\author[affiliation={1}]{Sujith}{P}
\author[affiliation={1}]{Visruth}{Sanka}
\author[affiliation={1}]{Nihar}{Desai}
\author[affiliation={2}]{Prasanta K}{Ghosh}

\address{
    $^1$ AI \& Robotics Technology Park (ARTPARK), I-Hub @ IISc, Bangalore, India \\
    $^2$Department of Electrical Engineering, Indian Institute of Science, Bangalore, India
}

\email{agneedh@artpark.in, pavanjk@artpark.in, sujith@artpark.in}


\usepackage{comment}
\usepackage{tabularx}
\usepackage{caption}
\begin{document}
\maketitle

% the abstract here must exactly match the abstract entered into the paper submission system
\begin{abstract}
    % 1000 characters. ASCII characters only. No citations.
Automatic Speech Recognition (ASR) systems are widely deployed across linguistically diverse regions, yet their ability to generalize across fine-grained geographic variation remains underexplored. We present a systematic study of cross-district ASR generalization for Indian languages, analyzing the impact of regional variation on performance. Using fine-tuning as a controlled probe, we train models on speech from a single district and evaluate them on other districts within the same language. We examine trends across multiple train–test district pairs and quantify performance differences. To assess geographic effects, we analyze the correlation between WER and inter-district distance using two distance measures. Our results show consistent correlations between geographic distance and WER, highlighting the challenges of regional generalization and the need for geographically diverse speech data in ASR development and evaluation in India.
\end{abstract}

\section{Introduction}

Automatic Speech Recognition (ASR) systems have become a foundational component of voice-driven technologies, enabling applications ranging from voice assistants and transcription services to accessibility tools. Despite rapid progress driven by large-scale pretraining and multilingual modeling, ASR performance continues to vary substantially across linguistic, dialectal, and regional populations \cite{Verma2023-rt}. This variability poses a significant challenge in linguistically diverse regions such as India, where speech exhibits pronounced variation even within the same language across relatively small geographic areas (\cite{bhanushali2022gramvaani}).

India is characterized by exceptional linguistic diversity, encompassing hundreds of languages and a far larger number of dialects and regional varieties \cite{censusindia2011}. The People’s Linguistic Survey of India reports approximately 780 languages \cite{devy2013peoples}, while subsequent studies estimate over 1,300 dialectal variations \cite{khanuja2021muril}. Recent work has further highlighted accent-level variability across dozens of districts \cite{lahaja2024}. These variations manifest as differences in phoneme realization, prosody, lexical choice, and speaking style, all of which can adversely affect ASR systems trained on pooled or standardized datasets. Consequently, ASR models that perform well on benchmark test sets often experience significant degradation in real-world deployments involving regional accents, sociolinguistic variation, or mismatched acoustic conditions \cite{FENG2024101567}.

Prior research has explored the impact of regional and dialectal diversity on ASR performance, particularly in the Indian context. Rao et al. demonstrated that regional accents substantially influence recognition accuracy and emphasized the importance of accent-aware modeling strategies \cite{inproceedings}. Similarly, Prakash et al. showed that fine-tuning ASR models on region-specific speech data leads to notable improvements in recognition accuracy for localized speech patterns \cite{kothari2023finetuning}. While these studies establish the effectiveness of local adaptation, they primarily focus on in-domain evaluation. As a result, a key question remains insufficiently explored: how well do ASR models adapted to one geographic region generalize to speech from nearby or distant regions within the same language?

In this work, we address this question by conducting a systematic, district-level analysis of ASR generalization across India. Using the Vaani dataset, which contains speech from over 156,000 speakers covering 165 districts \cite{vaani2025}, we study cross-district transfer in ASR performance. We use standard fine-tuning of a pretrained ASR model as a controlled experimental probe (\cite{singh2021multilingual}, \cite{javed2022indicsuperb}): models are fine-tuned using data from a single district and evaluated on speech from other districts within the same language. This setup allows us to quantify recognition performance degradation across a wide range of train–test district pairs. Beyond reporting word error rates, we analyze how ASR performance varies as a function of geographic separation between districts. By examining correlations between WER and multiple inter-district distance measures, we aim to understand whether geographic proximity serves as a meaningful proxy for linguistic and acoustic similarity at the district level. This district-centric perspective enables a finer-grained view of ASR robustness than prior state- or language-level analysis and reveals systematic patterns of regional generalization and failure.

To the best of our knowledge this is the first district-level quantitative study of ASR performance with distance in the Indian context. We give an empirical analysis of cross-district ASR generalization in a highly diverse linguistic setting. Our findings highlight the limitations of one-size-fits-all ASR models and underscore the importance of designing ASR systems with sensitivity to regional variation. These insights have practical implications for building more inclusive ASR technologies for real-world deployment in multilingual societies.

\section{Materials and Methods}

\subsection{Datasets}

The training datasets used in this study are sourced from Vaani. The Vaani dataset is a district-anchored dataset, which allows us to obtain sufficient data for each district-language pair.  Table 1 provides the duration (hrs.) of the dataset for each district. We selected these five languages and their corresponding districts to balance linguistic diversity, availability of high-quality training data, and geographic spread. The chosen languages represent different families within the Indo-Aryan and Dravidian groups, and covers both widely spoken and regionally concentrated languages. Within each language, the districts were chosen to reflect dialectal variation, while also ensuring sufficient training data availability from the Vaani corpus.

The test datasets were sourced from the Vaani Test Set, corresponding to the same district-language combinations as in Table~\ref{tab:dataset_details}. 
 
%%%%%%%%%%%%%%%%%%%%%%%%%%%%%%%%%%%%%%%%%%%%%%%%%%%%%%%%%%%%%%
\begin{table}[th]
  \centering
        \caption{Train Speech Dataset Details Across Districts}
        \label{tab:dataset_details}
  \begin{tabularx}{\columnwidth}{X l p{1.2cm} l }
    \toprule
    \textbf{District} & \textbf{Language} & \textbf{Dur.(hrs.)} & \textbf{State} \\
    \midrule
    Madhepura & Maithili & 2.25 & BR \\
    Samastipur & Maithili & 5.34  & BR \\
    Darbhanga & Maithili & 1.10  & BR \\
    Bhagalpur & Maithili & 0.52  & BR \\
    \midrule
    Bilaspur & Chhattisgarhi & 0.76  & CG \\
    Sarguja & Chhattisgarhi & 2.88  & CG \\
    Raigarh & Chhattisgarhi & 2.53  & CG \\
    Kabirdham & Chhattisgarhi & 1.21  & CG \\
    \midrule
    Mysore          & Kannada & 11.28  & KA \\
    Bellary         & Kannada & 10.16  & KA \\
    Dharwad         & Kannada & 9.73  & KA \\
    Dakshina Kannada & Kannada & 6.85  & KA \\
    Gulbarga        & Kannada & 5.06  & KA \\
    \midrule
    Jalpaiguri          & Bengali & 8.16 & WB \\
    Paschim Medinipur   & Bengali & 8.43  & WB \\
    Purulia             & Bengali & 8.67  & WB \\
    Kolkata             & Bengali & 11.8  & WB \\
    Dakshin Dinajpur    & Bengali & 9.65  & WB \\
    \midrule
    East Champaran  & Hindi & 7.00 & BR \\
    Muzaffarpur     & Hindi & 7.07 & BR \\
    Araria          & Hindi & 9.52 & BR \\
    Gaya            & Hindi & 6.81 & BR \\

    Muzaffarnagar    & Hindi & 10.93 & UP \\
    Etah             & Hindi & 10.68 & UP \\
    Hamirpur         & Hindi & 10.83 & UP \\

  \end{tabularx}

\end{table}
%%%%%%%%%%%%%%%%%%%%%%%%%%%%%%%%%%%%%%%%%%%%%%%%%%%%%%%%%%%%%%

\subsection{ASR Model and Adaptation Setup}
To examine whether geographic generalization trends are model-specific, we repeat all experiments using two architectures: Whisper\cite{whisper} and Wav2vec2\cite{wav2vec2}.
We use the \texttt{openai/whisper-small} model, and \texttt{facebook/wav2vec2-large-xlsr-53} as the base ASR systems. Whisper-small is a multilingual, 244M parameter transformer encoder-decoder based model pretrained on diverse speech data including those represented in our datasets. wav2vec2-large-xlsr-53 is a 300M parameter multilingual CTC-based model that is pretrained on 50+ languages including the ones in our dataset. The training and evaluation protocol is kept identical to ensure comparability.

To study cross-district generalization, we treat fine-tuning as a controlled experimental probe rather than a novel modeling contribution. For each language and district listed in Table~\ref{tab:dataset_details}, we fully fine-tune the models using speech data from a single district and evaluate them on speech from other districts within the same language. This setup enables direct comparison between in-district performance and cross-district generalization as a function of geographic distance. For each of the 25 districts in Table~\ref{tab:dataset_details}, we fine-tuned one whisper-small model and one wav2vec2-large-xlsr-53 model, totaling 50 models, spanning 5 languages, across 5 states of India.

Fine-tuning is performed by fully updating all model parameters. For each district-specific adaptation, the available data is split into training and validation sets in a 4:1 ratio. This uniform fine-tuning strategy ensures consistency across all experiments and provides a fair basis for analyzing how regional adaptation affects both in-district performance and cross-district generalization. By evaluating the baseline model alongside its district-adapted variants, we assess the extent to which standard fine-tuning improves recognition accuracy for localized speech and examine its limitations in handling dialectal and regional variation across geographic boundaries. This experimental design allows us to attribute observed performance differences primarily to regional linguistic and acoustic factors rather than model-specific or training-related differences.

\begin{table}
\caption{WER comparison between baseline and fine-tuned ASR models across languages.}
\label{tab:wer_language_comparison}
\centering

\begin{tabular}{lcccc}
\toprule
\textbf{Model} & \textbf{Language} & \textbf{Baseline} & \textbf{In-Dist} & \textbf{Cross-Dist} \\
\midrule
Whisper & Maithili      & 1.62 & 0.65 & 0.75 \\
Whisper & Chhattisgarhi & 1.45 & 0.54 & 0.61 \\
Whisper & Kannada       & 1.66 & 0.90 & 0.94 \\
Whisper & Bengali       & 3.07 & 0.50 & 0.51 \\
Whisper & Hindi         & 1.34 & 0.34 & 0.39 \\
Wav2Vec2 & Maithili      & - & 0.65 & 0.71 \\
Wav2Vec2 & Chhattisgarhi & - & 0.52 & 0.57 \\
Wav2Vec2 & Kannada       & - & 0.62 & 0.65 \\
Wav2Vec2 & Bengali       & - & 0.36 & 0.39 \\
Wav2Vec2 & Hindi         & - & 0.31 & 0.34 \\

\end{tabular}

\end{table}

\subsection{Evaluation}
We evaluate ASR performance using Word Error Rate (WER), the standard metric for speech recognition \cite{wer}. For each language and district, ASR models are evaluated on held-out test sets both before adaptation (zero-shot) and after district-specific fine-tuning. WER is computed for each train--test district pair to assess absolute recognition performance and cross-district generalization. Prior work on WER estimation uses Pearson correlation to quantify the agreement between predicted and actual WER values across speech recognition systems \cite{ewer2}. To quantify the relation between geographic distance and model performance, we compute Pearson’s correlation coefficient (r) between WER and inter-district distance. 

We consider two distance metrics to measure the geographic separation of two districts: (1) \textbf{Spherical Distance}: It is the shortest distance between two points on the surface of the Earth. The latitudes and longitudes of district centers are used and Haversine formula \cite{Sinnott1984} is applied to compute the distance. (2) \textbf{Adjacency Distance}: It is the distance of the shortest path between two nodes in a graph that is constructed by representing districts as nodes and placing an edge between two nodes if and only if the corresponding districts share a border.

\begin{table*}[t]
\caption{Mean Pearson correlation ($r$) between WER and geographic distance across models and distance metrics. $N$ denotes the number of districts per language used to compute fixed-condition correlations.}
\label{tab:distance_comparison}
\centering
\small
\begin{tabular}{llc cccc cccc}
\toprule
 &  &  & \multicolumn{4}{c}{\textbf{Whisper}} 
   & \multicolumn{4}{c}{\textbf{Wav2Vec2}} \\
\cmidrule(lr){4-7} \cmidrule(lr){8-11}
 &  &  & \multicolumn{2}{c}{\textbf{Spherical}} 
   & \multicolumn{2}{c}{\textbf{Adjacency}}
   & \multicolumn{2}{c}{\textbf{Spherical}} 
   & \multicolumn{2}{c}{\textbf{Adjacency}} \\
\cmidrule(lr){4-5} \cmidrule(lr){6-7}
\cmidrule(lr){8-9} \cmidrule(lr){10-11}
\textbf{Setting} & \textbf{Language} & \textbf{N}
& \textbf{Mean $r$} & \textbf{\% Sig.}
& \textbf{Mean $r$} & \textbf{\% Sig.}
& \textbf{Mean $r$} & \textbf{\% Sig.}
& \textbf{Mean $r$} & \textbf{\% Sig.} \\
\midrule

Train-fixed & Maithili      & 4
& 0.73 & 50
& 0.67 & 25
& 0.19 & 50
& 0.11 & 25 \\

Test-fixed  & Maithili      & 4
& 0.31 & 0
& 0.22 & 0
& 0.24 & 0
& 0.12 & 0 \\

\midrule

Train-fixed & Chhattisgarhi & 4
& 0.34 & 50
& 0.41 & 25
& 0.48 & 0
& 0.47 & 25 \\

Test-fixed  & Chhattisgarhi & 4
& 0.03 & 0
& 0.14 & 0
& 0.22 & 0
& 0.3 & 0 \\

\midrule

Train-fixed & Kannada       & 5
& 0.14 & 20
& 0.23 & 20
& 0.07 & 40
& 0.19 & 40 \\

Test-fixed  & Kannada       & 5
& 0.50 & 20
& 0.45 & 20
& 0.56 & 0
& 0.5 & 0 \\

\midrule

Train-fixed & Bengali       & 5
& -0.05 & 20
& -0.09 & 20
& 0.26 & 0
& 0.34 & 0 \\

Test-fixed  & Bengali       & 5
& 0.24 & 0
& 0.22 & 0
& 0.67 & 0
& 0.69 & 0 \\

\midrule

Train-fixed & Hindi         & 7
& 0.12 & 42.8
& 0.11 & 14.3
& 0.31 & 28.6
& 0.32 & 14.3 \\

Test-fixed  & Hindi         & 7
& 0.57 & 28.5
& 0.56 & 14.3
& 0.5 & 28.6
& 0.54 & 14.3 \\

\end{tabular}

\end{table*}

We consider the following two experimental perspectives for each of the distances mentioned before: (1) \textbf{Train-fixed Correlation}: The training district ($d_{\text{train}}$) is held constant, and we compute correlations as the test district ($d_{\text{test}}$) varies. $r_{\text{train}}(d_{\text{train}}) = \mathrm{corr} \left(
\mathrm{WER}(d_{\text{test}}, d_{\text{train}}),
    \mathrm{Dist}(d_{\text{test}}, d_{\text{train}})  \right)$.
    \textbf{Test-fixed Correlation}: The test district ($d_{\text{test}}$) is held constant, and we compute correlations as the training district ($d_{\text{train}}$) varies.
    $r_{\text{test}}(d_{\text{test}}) = \mathrm{corr} \left(
    \mathrm{WER}(d_{\text{test}}, d_{\text{train}}),
    \mathrm{Dist}(d_{\text{test}}, d_{\text{train}})
    \right)$.

\section{Results}

\subsection{Baseline vs Fine-Tuned WER}
Table~\ref{tab:wer_language_comparison} reports three evaluation settings: (i) Baseline, where the pretrained model is evaluated without fine-tuning (being an encoder only model, the Wav2Vec2 pretrained model is missing the baseline values); (ii) In-District Fine-Tuning (In-Dist.), where models are fine-tuned and evaluated on the same district; and (iii) Cross-District Fine-Tuning (Cross-Dist.), where models are fine-tuned on one district and evaluated on other districts within the same language. "Whisper" model refers to \texttt{openai/whisper-small}, "Wav2Vec2" refers to \texttt{facebook/wav2vec2-large-xlsr-53}. In-district fine-tuning yields the lowest error rates, while cross-district evaluation shows consistent increase in WER.

\subsection{Pooled Correlation}
\begin{figure}[t]
    \centering
    \vspace{0.5cm} 
    \includegraphics[width=\linewidth]{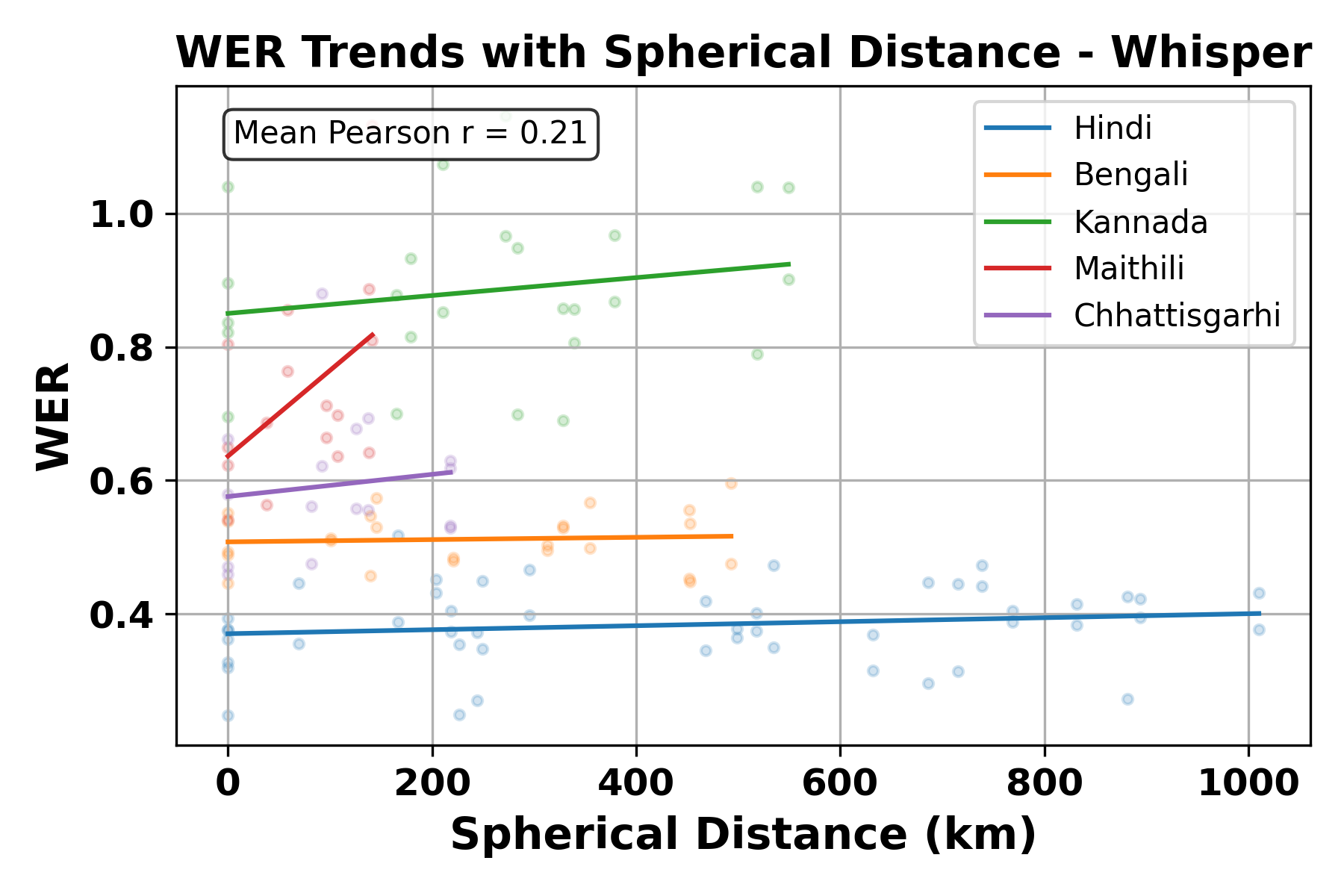}
    
    \vspace{0.7cm} 
    
    \includegraphics[width=\linewidth]{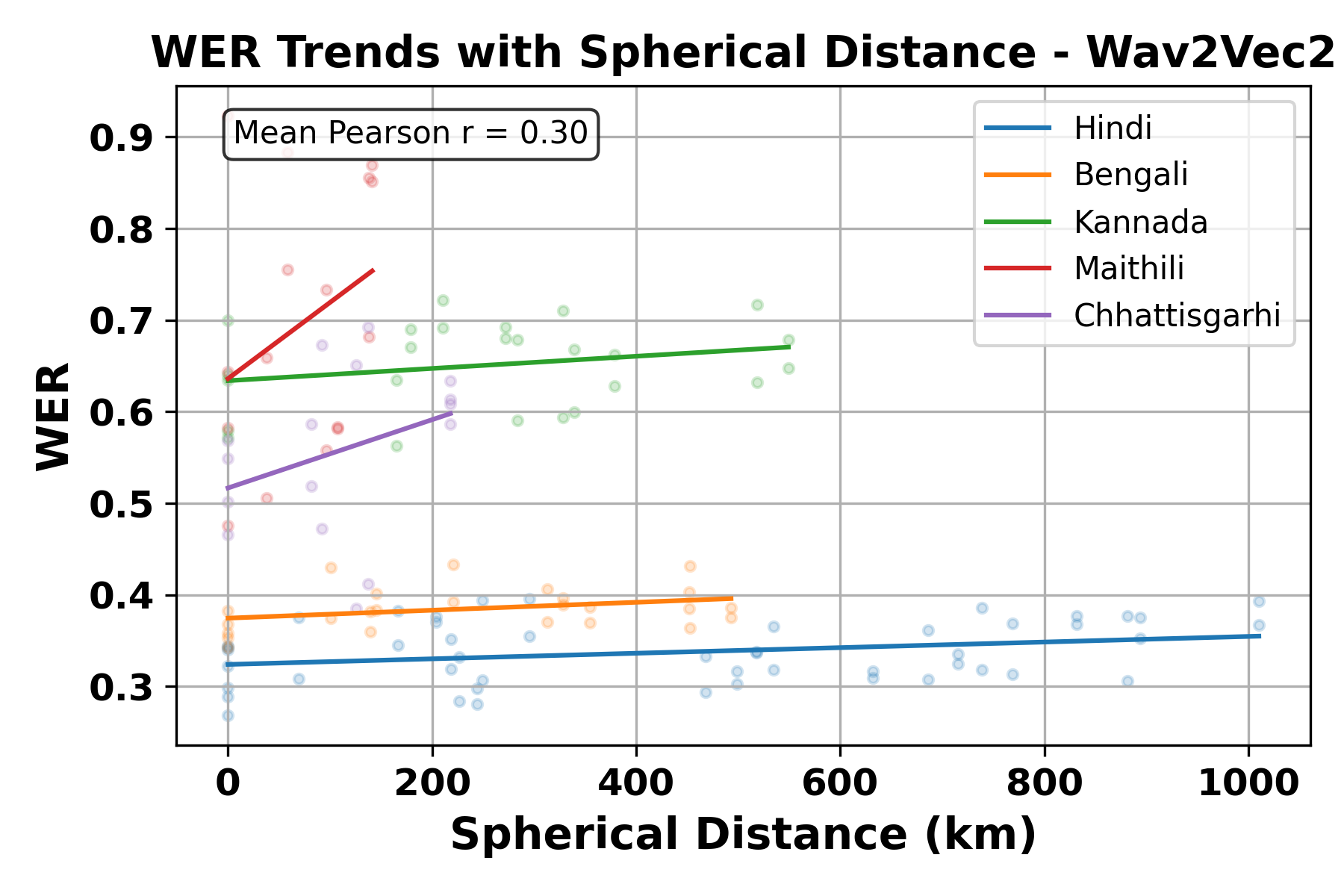}

    \vspace{0.7cm} 

    \includegraphics[width=\linewidth]{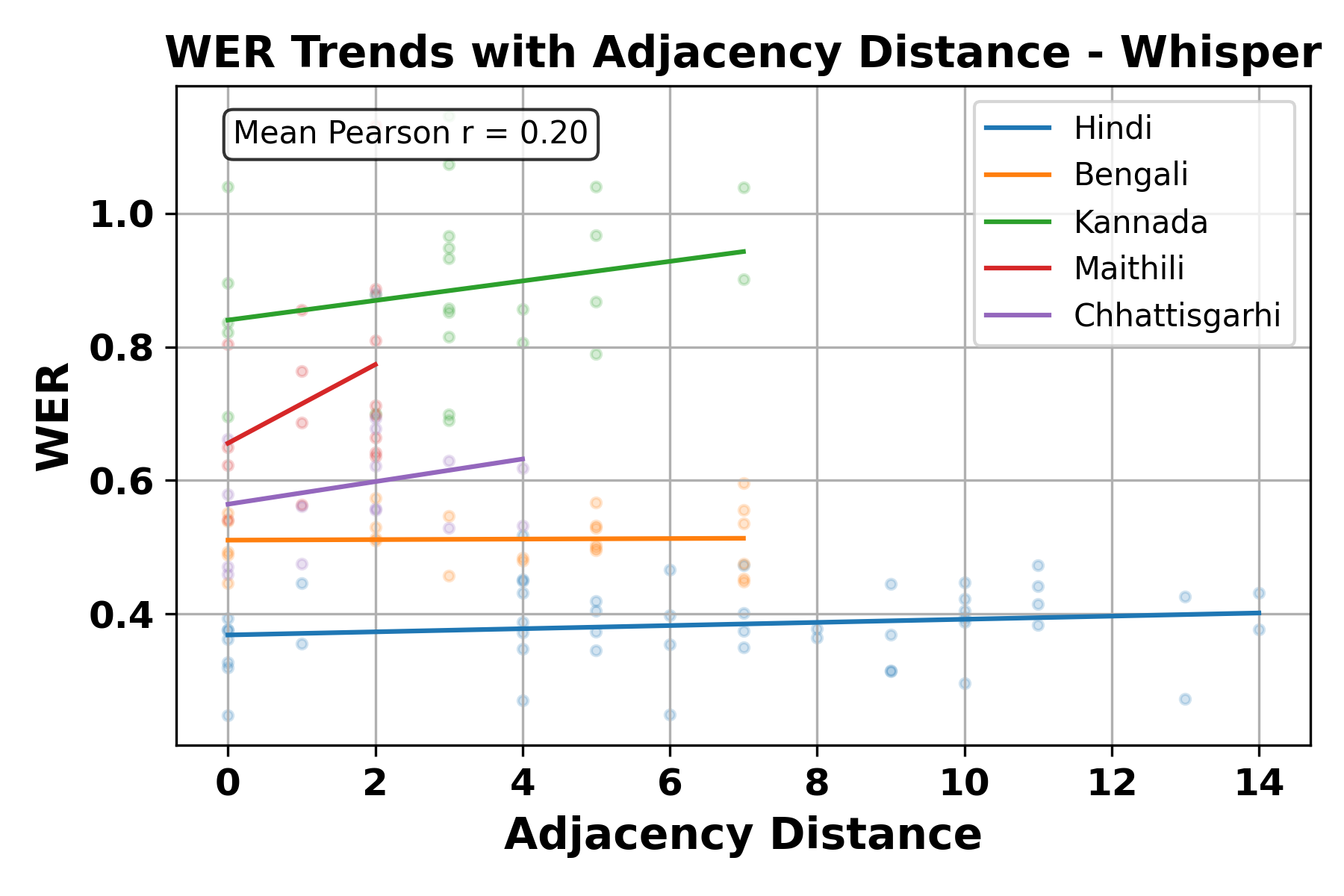}
    
     \vspace{0.7cm}
    
    \includegraphics[width=\linewidth]{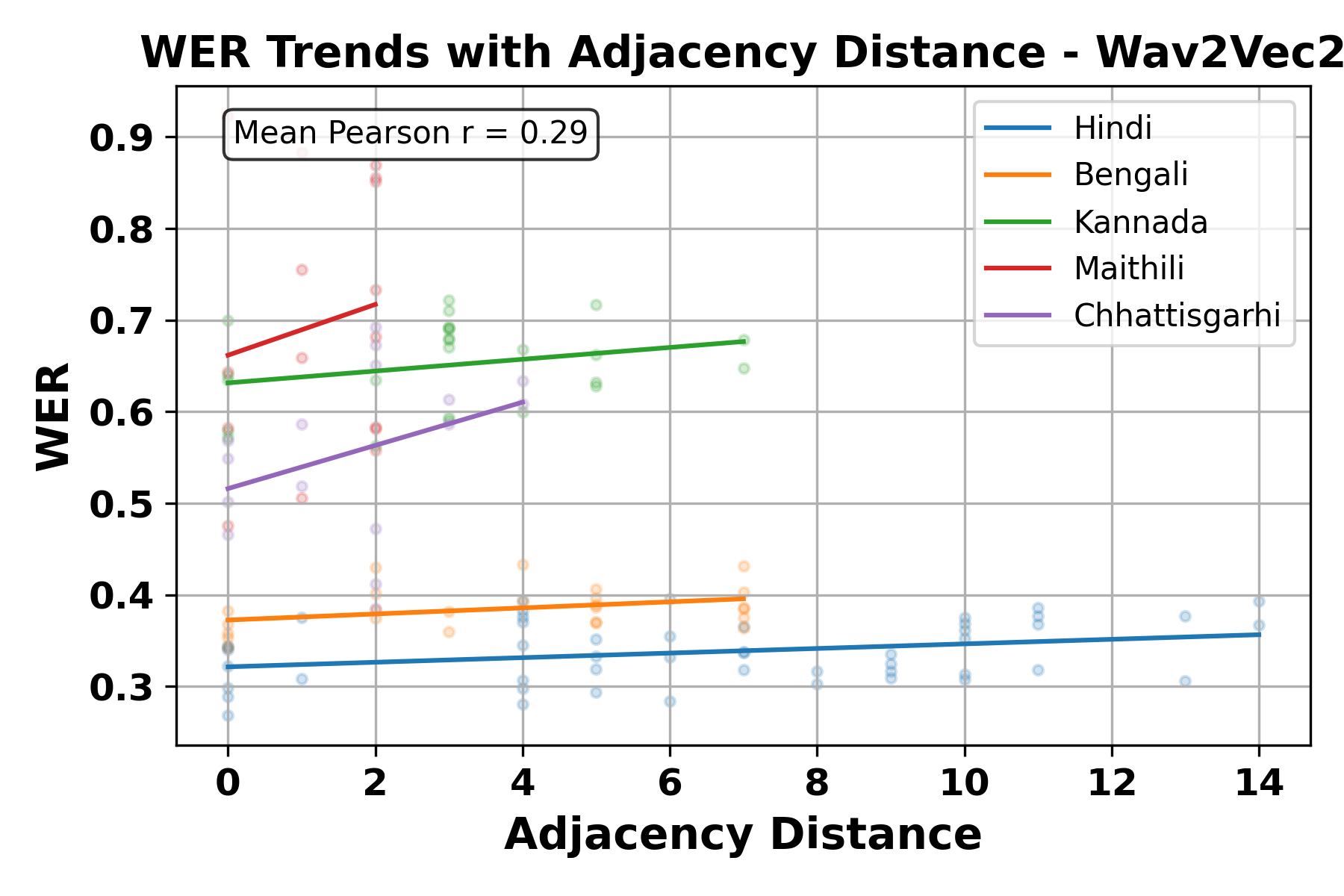}
    \caption{Pooled relationship between WER and distance across all train–test district pairs. Points represent individual train–test evaluations, and lines denote language-specific linear trends.}
    \label{fig:wer_trends}
\end{figure}

The pooled analysis considers all train–test district pairs jointly, without fixing either the training or test district. We examined the pooled relationship between WER and geographic distance across all train–test district pairs (Figure~\ref{fig:wer_trends}). Overall, there is a positive trend in WER with increasing distance (as seen in both distance measures, for both models), indicating that districts farther apart tend to exhibit higher error rates. While individual points show variability, the fitted lines suggest that this trend is generally consistent across languages with a mean Pearson correlation of 0.21 (for whisper-small) and 0.30 (for wav2vec2-large-xlsr-53) for spherical distance and 0.2 (for whisper-small) and 0.29 (for wav2vec2-large-xlsr-53) for adjacency distance, supporting the aggregated observation that geographic separation contributes to performance degradation.

\subsection{Correlation Summary}
Table~\ref{tab:distance_comparison} reports the mean correlation between WER and distance across models, languages, and distance metrics. Overall, the results support the hypothesis that distance from the training region is a consistent positive predictor of ASR error, though the strength of this relationship varies across languages and experimental conditions. Maithili and Hindi exhibit the strongest correlations (Whisper, r=0.73 and 0.57). Train- vs. test-fixed asymmetries suggest language-specific bias sources. Whisper leads for Hindi; Wav2Vec2 for Bengali (r=0.69). Spherical distance marginally outperforms adjacency, but both metrics agree directionally.

\subsection{Correlation Trend}
\begin{figure}[t]
    \centering
    \includegraphics[width=\linewidth]{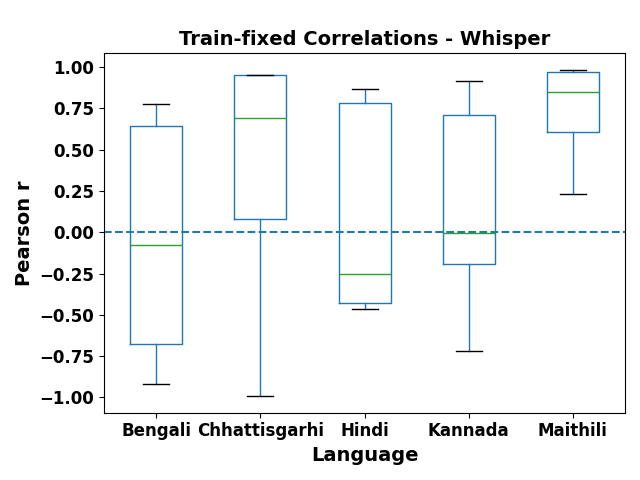}
    \includegraphics[width=\linewidth]{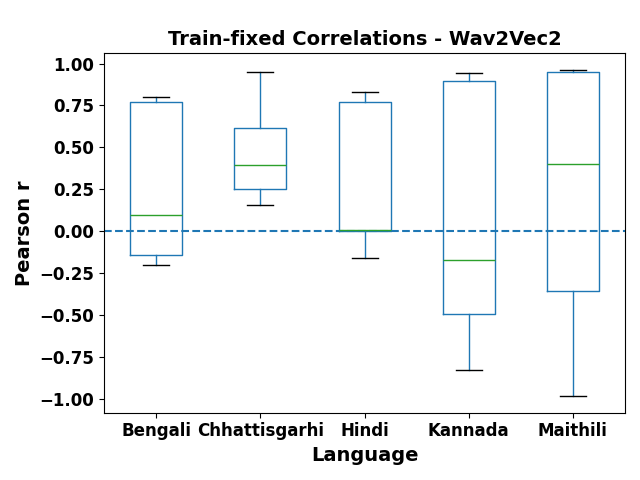}
        
    \includegraphics[width=\linewidth]{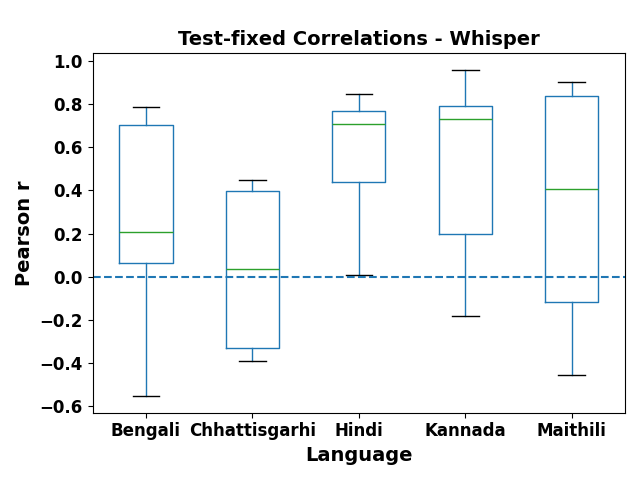}
    \includegraphics[width=\linewidth]{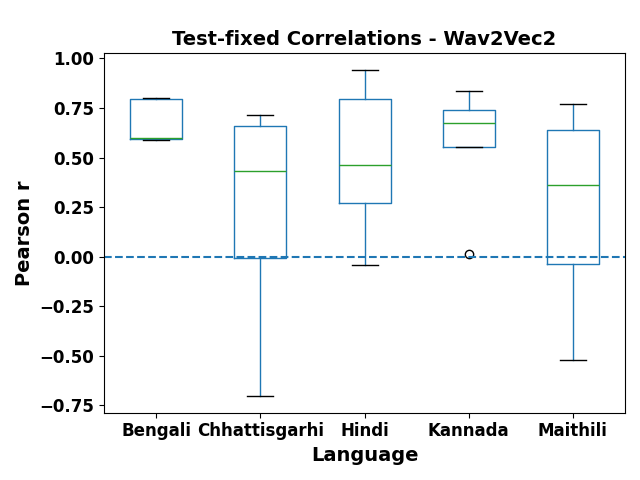}
    \caption{Train-fixed and Test-fixed Pearson correlations between WER and spherical distance across languages.}
    \label{fig:corr_hist}
\end{figure}

Figure~\ref{fig:corr_hist} visualizes the distribution of Pearson correlation coefficients between WER and spherical distance under the test-fixed and train-fixed settings. In both cases, the distributions are skewed toward positive values, indicating that greater geographic separation is generally associated with higher WER. This observation is consistent with the aggregated statistics reported in Table~\ref{tab:distance_comparison}. While individual correlations exhibit substantial variability due to the limited number of districts per setting, the overall trend remains stable across experimental settings. We treat correlations as descriptive effect sizes rather than hypothesis tests. This supports our focus on aggregated behavior rather than per-district significance-values.

%%%%%%%%%%%%%%%%%%%%%%%%%%%%%%%%%%%%%%%%%%%%%%%%%%%%%%%%

\section{Discussion and Conclusion}

We examine district-level ASR generalization in the context of Indic languages and find that although district-specific fine-tuning improves local performance, gains do not reliably transfer to distant regions, highlighting a trade-off between specialization and robustness. We also observe a general increase in WER as the distance between training and testing districts grows. This trend holds across languages, models, and evaluation settings, indicating that geographic proximity captures meaningful regional linguistic and acoustic variation. Train-fixed and test-fixed correlations are largely positive, reflecting regional and data asymmetries. Despite small sample sizes in some cases, aggregated cross-language results reveal a stable directional effect.

Overall, these findings provide empirical evidence that geographic factors influence ASR performance in the Indian context and underscore the need for geographically informed evaluation and region-aware adaptation strategies for robust deployment in diverse settings.

\clearpage

\bibliographystyle{IEEEtran}
\bibliography{mybib}

\end{document}